# Patent Mining by Extracting Functional Analysis Information Modelled As Graph Structure: A Patent Knowledge-base Collaborative Building Approach

*Manal E. Helal[1], Mohammed E. Helal[2]*

[1]School of Physics, Engineering and Computer Science, University of Hertfordshire, Hatfield, UK , m.helal@herts.ac.uk

[2]Computer Engineering Department, AASTMT, Cairo, Egypt , mdehelal@gmail.com

## Abstract

Patents provide a rich source of information about design innovations. Patent mining techniques employ various technologies, such as text mining, machine learning, natural language processing, and ontology-building techniques. An automated graph data modelling method is proposed for extracting functional representations for building a semantic database of patents of mechanical designs. The method has several benefits: The schema-free characteristic of the proposed graph modelling enables the ontology it is based on to evolve and generalise to upper ontologies across technology domains and to specify lower ontologies to more specific domains. Graph modelling benefits from enhanced performance of deep queries across many levels of relationships and interactions and provides efficient storage. Graph modelling also enables visualisation libraries to use the graph data structure immediately, avoiding the need for graph extraction programs from relational databases. Patent/Design comparisons are computed by search queries using counting of overlaps of different levels and weights.

This work has produced the PatMine SolidWorks Add-in ©, which compares annotated CAD designs with patents and highlights overlapping design concepts. The patent annotation extracts its functional analysis, representing its structure as geometric feature interactions. Additional features such as full-text search and semantic search of the PatMine patents database are available, and graph analytic methods and machine learning algorithms are enabled and can be implemented as plug-ins in future work.

**Keywords:** Patent Mining; Semantic Analysis; Functional Analysis Diagrams; Graph Data Modelling; Visualisation; Similarity Scoring; Big Data Analytics; Machine Learning; Artificial Intelligence; Natural Language Processing

## Introduction

Patent filing, applications, storage and archival present a big data problem that requires automated mining for prior art knowledge discovery and understanding. Patent mining is a new research direction to distinguish it from other data and text mining approaches. The pre-filing mining process requires classifying the patent documents, searching for references, and deciding whether to accept the patent based on innovation scoring. Post-filing mining is required to decide whether to maintain the accepted patent and whether there is any infringement of existing patents. These analytical tasks are mainly: 1) classification (categorising patents into classes and subclasses); 2) retrieval (which patents come up in a search query using various query objectives); 3) visualisation (summarising patent information graphically); and 4) valuation (innovation scoring, infringement decisions) (Zhang et al., 2015)

A patent document contains structured information such as patent number, inventor, filing date, issued date, and assignees. It also contains unstructured information of varying lengths, such as abstract, independent and dependent claims, invention descriptions, detailed specifications, and various drawings to illustrate the solution idea. Although more structured than a typical web page, patents are usually longer, contain deep technical and legal terms and might be written in different languages. This makes the successful keywords full-text search used for web pages not very successful with a patent search, rendering patent analysis to be more difficult even for domain experts (Zhang et al., 2015). The knowledge in engineering patent documents has been systematically extracted using various methods such as Quality Function Deployment (QFD), Value Analysis (VA), Theory of Constraints (TOC), and Axiomatic Design (AD). The CTOC method (Converter, Transmitter, Operator, Control/Command) has been used to enhance the choice of keywords through physical analysis of the patents' concepts, employing a database of energy converters and the evolution trends of technical systems (Valverde et al., 2017). Ontologies are defined to use controlled vocabulary, taxonomy, glossaries, and semantic relationships to systematically extract structured semantic information to store in semantic databases for further queries and analysis tasks. The use of ontologies is common in patent informatics research, as will be discussed in the methods sections.

This paper proposes a novel method of combining ontologies and semantic databases in a graph data model to capture the functional analysis diagrams of patents and store them in a patent knowledge base. This method has various performance and analytical benefits, visualisation readiness, and enables future technologies to be embraced more readily than alternative methods.

A case study of examples is presented that according to the International Patent Classification (IPC) taxonomy, the example patents are classified under the (B) "Performing operations, transporting". The Cooperative Patent Classification (CPC) class of interest is CPC Subclass B65D, which is concerned with containers for storing or transporting articles or materials. The study aims to: 1) create a semantic database of information extracted from patents by choosing the most suitable data model; 2) visualise patent contents in informative illustrations; 3) and enable scoring of similarity and other patent analytics objectives.

The patent files in CPC Subclass B65D are usually supported by engineering drawings describing the design's key aspects. For many mechanical engineering designs, the innovative working principle in the functional claim is achieved through the design's geometric components and their functional interactions. Reconciled Functional Basis (RFB) (Hirtz J. et al., 2002) was introduced as a controlled vocabulary knowledge base in which known function verbs, flows, or objects of actions are arranged hierarchically. There were attempts to capture the high connectivity of functional interactions and enrich the knowledge base by adding physical, chemical, and logical levels (Otto and Wood, 2001). Capturing these interactions is achieved through various methods in the literature: Function trees (*Value Analysis Incorporated (VAI)*, 1993), Data Flow Diagrams (Yourdon E., 1989), and Function Structure (Pahl G. et al., 2007) are form-independent representation methods (not co-related to the geometric components of the design) to describe the overall function and sub-functions of the design. A form-dependant functional modelling method was introduced as Functional Analysis Diagrams (FAD) (Aurisicchio et al., 2012), which is a graphical mapping of the highly interconnected mesh of all interactions and flows of energy, data, and events of an engineering device. FAD is dependent on the device's parts (geometric features) and can be drawn on top of the product's engineering drawing.

The methods in this study are developed by attempting various approaches from the literature that are explained in this paper as a survey of these technologies and when they are more useful. The methods section is divided into four subsections. The first subsection describes the data models literature and which approach is more suitable to the problem addressed in this study. The second subsection explains the semantic modelling of databases literature. Both subsections introduce graph modelling as a solution to the large and complex data problems of patents and the required semantic modelling and visualisation of its components. The third subsection briefly describes how working principles of mechanical engineering patents can be modelled using Functional Activity Diagrams (FAD). The fourth subsection demonstrates the contributions of the proposed graph data modelling method by describing how FAD models for patents and emerging designs can be modelled using graph data structures and how this enhances the performance of many of the patent mining tasks discussed in the introduction. The results section discusses the experimental results and the production of the PatMine SolidWorks Add-in © plus its current development status and possible future directions to enhance its performance. The results section is divided into four subsections. The first subsection describes the user interface through which the user enters the FAD model for a given patent or emerging design. The tool generates the cypher statements queries to store the FAD model in a graph database for future search, computes similarity scores and generates the graph visualisation to enable further analysis of the model intuitively using clustering and groupings of interactions. The second results subsection explains the various search methods; each will generate a different cypher query and affect which patents appear in the search results. The third results subsection explains the similarity scoring method. The fourth results subsection discusses the patents and graphs visualisation literature in relation to what is employed in this study. Finally, conclusions are drawn.

# Methods

### *Data Models Literature*

Since the 1970s, various data models have been proposed in the literature to represent, store, maintain and query abstractions of real-world entities. A data model defines a set of data structures (mathematical frameworks to represent knowledge), a set of operators or inference rules, and integrity rules (Simsion and Witt, 2005). The earliest models were focused on actual file system organisations using either hierarchical or network models. Later, the relational databases model was proposed to offer a separation of the logical and physical models. This led to the entity-relationship model describing the conceptual data model from the user point of view to be independent of the Relational DB Management Systems (RDBMS) that manages the physical data model and how it provides adequate performance and distributed access mechanisms. Then Object Oriented (OO) models were introduced in the 1980s to capture the hierarchical organisation of data structures and reduce their complexity by using inheritance (Kim, 1990). RDBMS and OO data models place all data as tuples representing relations (entities mapped to their values of particular attributes) in rows with headers identifying column names (attribute names) in tables similar to spreadsheets (Angles and Gutierrez, 2008).

Motivated by the lack of semantics in RDBMS, Graph Data models were proposed in the 1990s to emphasise the interconnectivity of entities and offer unstructured, semi-structured and structured organisation of data elements. The database research communities shifted their attention to the semantic web giving rise to semantic databases using information exchange models mainly designed for the World Wide Web, using technologies such as Extensible Mark-up Language (XML), Resource Description Framework (RDF) and Web Ontology Language (OWL). These models can use hierarchical tree-like structures, networks, linear, or application-specific data structures. They are primarily built on top of relational databases or others for storage (Angles and Gutierrez, 2008).

The limitations of RDBMS shifted the research directions to NoSQL databases (Not only SQL) to satisfy the requirements of Big Data distributed storage, manipulations and retrieval scaling and performance requirements. These are categorised as four models: 1) wide column store approaches such as Google's Big Table model, Apache Cassandra, and Amazon Dynamo; these are no-schema extensible stores in which the two-dimensional key-value tables have the ability to extend to more dynamic tables; 2) Document Stores for semi-structured data such as MongoDB; 3) Key-value stores such as BerkelyDB; and 4) Graph Databases which uses graph-like data structures (Angles, 2012).

The graph data structures can be: 1) simple graphs (just a set of nodes connected by a set of edges); 2) hyper-graphs consisting of hyper-nodes and hyper-edges; hyper-nodes (nested graphs) offer nesting of nodes, such that one node can contain a subgraph of other nodes and their relations; hyper-edges offer relations that connect more than one node; 3) attributed graphs (contain properties describing nodes and edges); 4) directed edges (unidirectional) or undirected edges (bidirectional); 5) labelled or unlabelled (typed nodes and edges or untyped). Implementation details vary across the graph DB systems on whether to support hyper-graphs, directions, attributes and labels and how these features are implemented (Angles and Gutierrez, 2008).

Graph data models have evolved from semantic networks for meta-data descriptors of databases to logical data models that generalise the relational, hierarchical and network models. The following Graph Modelling Systems are introduced in the literature, each with different features and advantages. G-Base represented complex structures. $O_2$ used graphs to represent Objects. GOOD is a Graph Oriented Object Model. Then GMOD, GRAM, and PaMaL included tuples and sets. GOAL introduced association nodes. G-Log is a declarative language for graphs. GDM introduced n-ary symmetric relationships. There are also a Hyper-node model, multi-scaled networks, and GROOVY as an object-oriented graph model that uses hyper-graphs. GraphDB modelled transport networks in an Object Oriented database. Graph Views used either relational Object Oriented database systems or file systems to store graphs. GRAS models complex information from software engineering projects using attributed graphs. OEM focuses on information exchange from heterogeneous information sources (Angles and Gutierrez, 2008).

Initial Graph Query Languages are based on the relational tuple calculus, upon which SQL is based. G, G+, and GraphLog are Graph Query Languages defined over a general simple graph model that elaborates a regular expression-based query to include a summary graph to restructure the answer. G-Log is a graph-based, declarative language, non-deterministic and computationally complete graph query language. Graphical node insertions and deletions (graph transformations) as means of database transformation and pattern matching are offered in GOOD, GMOD, PaMaL, GOAL, GUL. Later on, recursion, loop procedures and program constructs were included. Hyper-graph Manipulation Language (HML) is introduced to handle hyper-graphs querying by identifiers or values and addition/deletions operations for hyper-graphs and hyper-edges. Gram and GoQL are SQL-like graph query languages that offer path expressions. DGV and GraphDB perform several steps to extend or restrict the resulting graph dynamically. Path problems provide solutions to many graph queries, such as reachability and neighbourhood degree, centrality and various forms of connectivity and pattern matching. G-Store and Sones include SQL-based query languages. Other graph query language models were proposed in the literature advancing their functionalities over time and targeting various applications (Angles and Gutierrez, 2008).

The recent graph query languages include SparQL as an RDF Query language that consists of triplet patterns, conjunctions, disjunctions and optional patterns. SparQL lacks various essential constructs, such as looping and branching. The recent choice of graph query languages is limited to GraphQL, Gremlin, and OpenCypher. GraphQL is the data query language used in Facebook and is not the most comprehensive QL for graphs. Gremlin is a graph traversal language by the Apache project and was adopted by Neo4j and OrientDB. Cypher is proposed as the Neo4j Graph Query language. OpenCypher is an open-source QL based on Cypher (Panzarino, 2014).

Graph data modelling is preferred when the interconnections or the topology of the data entities are more or equally important than the data entities themselves. Graphs are most suitable when not all values are present and/or the schema of the dataset is not known or known to change over time. Also, this model is preferred when data is big, and visualising it in tables will not convey much information.

When modelling data using graph data structures, the degree of separation between schema and data needs to be decided by the application's requirements to specify the integrity constraints to enforce. Initial graph systems are used to model nodes in a tree whose internal nodes are structured data, and leaves are the data. Another structure for graphs used two-column tables. Schema-based graph DB models are very restrictive. Table-based graph DB models perform slower than physical graph models in which the database file system contains a list of relationship records. A graph physical file system offers the simplicity of no-fixed schema, only defined nodes and relations, ordered by their defined labels (type) and the direction of the relationship. When properties and their values exist, they are stored. Indexing these files will speed up the traversal of relationships to be much faster than indexing attributes in table-based file systems and their SQL Join queries (Gupta, 2015).

Graph DB is manipulated using graph operations as graph transformations. The graph operations are mainly: adjacency, reachability (path problems), pattern matching (isomorphism of sub-graphs), and summarisation queries (aggregation and statistics). In Social networks, the graph operations are mainly: distance, neighbourhoods, paths, subgraphs, graph patterns, connectivity, graph statistics such as diameter and centrality, clustering coefficient of a vertex, clustering coefficient of a network, betweenness and size of giant connected components, and size distribution of finite connected components. In Geographic Information Systems (GIS), the graph operations are mainly: 1) geometric operations such as area or boundary, intersection, and inclusions; 2) topological operations such as connectedness, paths, and neighbours, and 3) metric operations such as distance between entities, the diameter of the network. To query the Database, natural language query language is used (Angles and Gutierrez, 2008).

The recent attention given to graph databases led to the development of new Graph database systems such as Neo4j, AllegroGraph, HyperGraphDB, DEX, InfiniteGraph, Sones, InfoGrid, Sones and VertexDB. The comparative study in (Angles, 2012) studied these graph databases in terms of data structures, query language and integrity constraints. The benchmarking on HPC Scalable Graph Analysis of Neo4j, Jena, HypergraphDB and DEX graph databases in (D. Dominguez-Sal et al., 2010) concluded that DEX and Neo4j are the most efficient implementations. The empirical study in (Jouili and Vansteenberghe, 2013) compared Neo4j, OrientDB, Titan (BerkeleyDB and Cassandra) and DEX graph databases to conclude that Neo4j outperforms all the other candidates, regardless the workload or the parameters used.

Neo4j community edition was selected as the graph database system for the solution proposed in this study. Neo4j is a schema-free graph database system. Neo4j supports main memory and external memory storage schemes to handle large data. Neo4j implements indices to improve data retrieval operations and uses APIs to handle data operation and manipulation. It offers attributed graphs (attributed nodes and edges), labelled/typed nodes and edges, and directed graphs (Gupta, 2015).

Multiple labels for nodes and edges in graphs offer opportunities to include different layers of abstractions, such as Object Oriented inheritance and "is-A" relationships (typing). Properties or attributes offer the simplicity of describing the node or edge in isolation or in relation to its context. Although Neo4j does not support hyper-graphs and multilevel explicitly, using multiple properties or as keys or edges to retrieve other sub-graphs. The same can be used to implement multiple connections to offer the hyper-graph functionalities and "is-Part-of" relations such as aggregation, composition, association, sets, and n-ary.

The object identity constraints/type checking cannot be enforced in Neo4j, as labels are defined by users as per the data specification requirements or the application requirements. Entity Integrity can be constrained by adding a unique constraint on a particular property for nodes of particular labels. Duplicates can be avoided by merging instead of creating. Referential integrity can be constrained in every newly created link by matching before linking. There is no schema constraint to enforce referential integrity. Also, semantic integrity (such as value functional dependencies) can be enforced using scripts to check for application-based constraints. Cardinality checking and graph pattern constraints can be added programmatically from the application rather than the Neo4j server backend.

## *Semantic Databases*

CAD/CAM software is a data-intensive domain that involves complex data objects and complex object interactions that require a knowledge base rather than traditional relational database models. Object-Oriented (OO) database models were first used to address this complexity. OO models require fixed schema and inheritance structure but allow for semantic context. Both Semantic and OO data models have graph data structures embedded in them. There is research interest in resolving the limitations of these models to address the requirements of AI algorithms. This direction enables the realistic representation of complex real-life relationships such as in social, information, technological, and biological networks (Angles and Gutierrez, 2008).

The Integrated Computer-Aided Manufacturing (ICAM) Program, aimed at increasing manufacturing productivity, defined the need for better analysis and communication techniques using computer technologies known as IDEF (ICAM Definition) for formal functional modelling (*ICAM Conceptual Design for Computer Integrated Manufacturing Framework Document*, 1984). IDEF0 is the functional model, IDEF1 is the information model, IDEF1x is the semantic model, and IDEF2 is the dynamic model. There are similar projects such as the "Product Data Exchange Specification" (PDES) project in the US, the related ISO Standard for the exchange of product model data (STEP), the Computer Integrated Manufacture Open Systems Architecture (CIMOSA), and [ISO87] project in the European Economic Community. There is also ISO 15926 as a standard for data integration, sharing, exchange, and handover between computer systems. These projects enable useful data sharing by using formal semantic data models in the context of the data described (Edward J. Barkmeyer, 1989). ISO 18629 is a process specification language (PSL) to share computer interoperable statements based on first-order logic and situational calculus serving as the foundational theory (Voth, 2005). These projects emphasise the importance of clarifying the true semantics of computerised CAD/CAM models, whether for manufacturing precision improvement and automating or for knowledge sharing and innovation evaluation.

Semantic models started in the early 1970s as schema design tools. Away from the record-based models, Semantic models give relational data meanings from the viewpoint of the user using Artificial Intelligence (AI). Semantic models are a sort of standard abstraction for knowledge representation that enables automated reasoning and machine learning in AI algorithms. Semantic models are high-level abstractions of the rich and more complex real-life semantics that are required to be processed in a natural language. This is achieved by modelling the data relationships arising from typical relational Databases in the third normal form, creating a more complex navigation of data relationships in an Object Oriented paradigm style.

Semantic DB models or management systems (SDBMs) increase the separation of physical and logical models; they offer more constructs for describing data interrelationships, moving some schema information from the constraints to the structure side; they also offer several abstraction mechanisms such as levels of details to view, modularity degree, and derived schema components. The schema is controlled and represented in traditional DBMS (Hull and King, 1987).

The initial semantic models included the structural model that identified five types of relations from the relational model. IFO, RM/T, GEM, and Galileo identified relationships and updated semantics as extensions to the entity-relationship (ER) model. Early SDBMS used existing DBMS to build efficient data access mechanisms such as EFDM, TAXIS, ADAPLEX, and Sembase. Handling the dynamics of the semantic models, TAXIS and Galileo synthesise semantic modelling constructs using control and typing mechanisms from imperative programming languages. On the other hand, the Extended Semantic Hierarchy Model (SHM+) and INSYDE develop control mechanisms that follow the structure of the semantic schema.

Semantic Query Languages initially included DAPLEX, GEM, and ARIEL, which extended the ER query language based on relational calculus. FQL is Functional Query Language based on functional programming. FQL does not include schema update definitions. TAXIS, DIAL, Semdal, and Galileo, are imperative languages with the standard flow of control facilities and arithmetic capabilities

Gellish is another example of a semantic Database that uses the Gellish Formal languages, such as English Gellish, as a controlled language to conceptualise computer-interpretable facts (Van Renssen, 2005). This sort of Natural Language Processing (NLP) using language dictionaries and taxonomy is different from OWL, which does not use dictionaries or depend on a particular language. OWL is a family of knowledge representation frameworks to create domain ontologies from structured corporate databases. OWL is also extensible and creates a form of upper ontology that is general to all domains. OWL 2 has been the formal language for semantic ontology editors such as Protégé. OWL is built upon the W3C XML standard objects called Resource Description Frameworks (RDF). RDF is a conceptual data modelling approach that is suitable for the semantic World Wide Web and its knowledge management applications. RDF uses key-value pairs, or triplets of subject-predicate-object, or maybe Quad-stores if named graphs are used. RDB to RDF standards is published by W3C (Sahoo et al., 2009). The work in (Angles and Gutierrez, 2005) studied RDF from a graph database perspective and identified graph support in the RDF query languages and presented desirable graph primitives to incorporate in them, such as paths and connectedness, neighbourhood, and pattern matching.

The operations of a semantic database are more structural and hierarchical such as in Object Oriented models, in which inheritance is the key relationship. These operations are summarised as aggregations, classification, instantiation, sub and super–classing, and attribute inheritance. The triplets in RDF form a network structure in which the predicate is the edge describing a relationship between two nodes (entities). Both semantic models and graph models reason about the graph-like structure generated by the relationships between the entities. Semantic DB is different from graphs essentially in the schema requirements for the former and the schema-free nature of the latter. Another difference is that RDF imposes an index to traverse the resources, while graph databases support index-free adjacency and provide inferencing capabilities. However, later RDF versions cease to impose indices (Vyawahare Harsha and P. P. Karde, 2015).

Ontology building is a means of controlling the taxonomy that describes the formal naming and definition of the types, properties, and interrelationships of the entities in a given domain of knowledge. They can use any language but can be language specific such as WordNet, which is a lexical database for the English language (Fellbaum, 1998). Ontologies can be used in semantic information extraction from databases or documents. Ontology research deals with semantic heterogeneity in structured data. Handling a varying schema for unstructured data is not best handled using an ontology because not only will more semantics be added to the ontology gradually, but existing semantics might need to evolve and change over time.

## *Functional Modelling of Patents*

A functional description of the working principles in the patents and the emerging design is required for the assessment of possible conflict with prior art or the level of innovation. Conflict with the prior art is based on patent claims, and innovation can be a change in functional target, a new scientific effect principle, or an improved functional structure (Li et al., 2017). Frequency-based keyword search is not very useful for identifying these aspects. Recent research (Li et al., 2017) uses a functional ontology model to correlate the semantic extent of keywords in various design domains, then uses a co-reference relations library to decide the term weight and its association with the functional groups, and finally builds a naïve Bayes classifier using a training dataset of patents.

The FAD method (Aurisicchio et al., 2012) was proposed to simplify form-dependent functional description using an intuitive graphical notation where blocks represent device structure or other resources, and labelled arrows represent functions and information flows. Thus FAD captures the highly interconnected mesh of functional interactions within a system, enabling the representation of hierarchical structures and sub-functions. FAD can be laid out on the actual positions of the geometric components explaining the rationale of the design and enabling design improvements. FAD links product structure with functions and behaviours in natural language instead of the controlled RFB ontology with its limitations (Aurisicchio et al., 2012).

Geometric feature details are required in order to distinguish between mechanical engineering designs where function depends upon interactions between certain geometric features. A possible direction is to model the FAD model annotation as a triple-store using tools such as Protégé ontology editor to guide what to look for (taxonomy) in a patent document using RFB as an initial upper ontology. RFB was criticised in the literature as not covering all required concepts, types, interactions and all required entities, which is why an evolving dynamic schema and glossaries and their meanings are required. The taxonomy, glossaries, and ontology are currently designed in a structured manner to be approved by the particular design domain experts. The choice of a structured semantic database for the FAD models has several drawbacks, as follows:–

First: ontologies use structured fixed schema. This limits the evolution of the structure of interactions to the schema employed and the terms approved by the particular group of domain experts that designed the ontology initially.

Second: semantic database visualisation options are very limited in the literature. This reduces the FAD visualisation automation options to tailored visualisation software that takes a long time to develop for a particular schema. Alternatively, visualisation can be manually achieved by using third-party software packages such as designVUE (Imperial College London, 2016), a branch of VUE (Visual Understanding Environment) originally developed by Tufts University.

Third: semantic databases are usually built on top of relational databases, in which search performance is dramatically affected by the depth of the search query. Searching an interaction between two geometries will be in the order of the number of geometries defined in the Database, while searching a three level interaction will degrade the performance, and so forth. Using indices will enhance the performance. However, the depth of the query will always be a factor in the performance deterioration.

Fourth: the complexity of the FAD model resulted in limitations in quantifying the similarity between two complex FAD models. Simplification of the model is desirable, but not at the expense of losing details that can differentiate one FAD model from another. Conceptual comparisons and variable penalties of insertions and deletions to networks allow the class of network alignment algorithms to produce a more accurate score.

## *Graph Modelling of FAD/RFB Ontology and Patents Semantic Database*

This paper proposes the graph modelling of FAD models to capture the basic geometric features interactions and their functional behaviours, termed Functional Geometry Interactions (FGIs). FGIs use the RFB for the fundamental functional interactions to be defined by the designer. FGIs also capture the topological relationships between geometric features. The novel working principles of the design are captured as Key-FGIs. FAD is modelled in three levels: First, the abstract level captures the functional design claims in natural language. The second level is the detailed sub-components, such as the geometric features it contains. The third level captures the associated structure of the functions claimed and how they are achieved using the geometric features interactions.

As FAD is basically a set of FGIs, this enables an intuitive graph model in which the basic node type is the geometric feature, and the basic relationships among the geometries are the functional interaction edge. All other information is modelled as properties of the nodes or the edges.

The basic definition of a relation is to consider every property value that can be an entity of its own as such, and create relations between the main entity and the property entity, which is how Object Oriented Programming works (Bordoloi and Kalita, 2013). Analysis of the statistical use of a node value as a property to other nodes will be visual, but the visualisation of the graph will contain all these properties as separate nodes with edges connecting them and will be very complicated. Modelling the dataset with the visualisation objective in mind requires using labels and properties to hide the complexity and use only node types required to be visible in the graphical results. The hidden complexity in properties' values is available for retrieval using queries for statistical analysis.

A given FAD model was modelled in the Neo4j schema-free graph database system. The FAD model can describe a design proposed in a patent document or an emerging design modelled in CAD software that is being checked for similarity against other FAD models (for existing patents or designs in general) in the PatMine database. The glossaries and synonyms of the domain values acceptable for the various properties types, such as from the RFB ontology, are uploaded to the system embedded in the PatMine SolidWorks Add-in © User Interface (UI) as pre-filled selection options for the user to choose from. For example, geometric types (labels) and action values (FGI action property values). These values are loaded from external lookup tables in CSV formats to be used in defining the FAD model or while constructing search queries. For example, a query can include (geometry = "cuboid" or geometry = "block"). The user can add more user-defined options by editing the combo boxes UI controls. Only properties given values by the user are stored in the Database or used in the generated cypher queries.

From the FAD models, objects are mostly considered geometric features. Therefore the main node type is the geometry type (labelled as geometry). It represents the list of geometric objects included in a 3D engineering drawing (or CAD model). All attributes describing it are properties of the geometry node type. Each patent will be represented graphically as a number of nodes (circles) corresponding to its geometric object components. If a design contains more than one product, then each will have its own subgraph of geometrical nodes. The PatMine geometric features levels of abstractions for the hierarchical is-A relationship are modelled as extra labels. For example, an object in level 5 is a square sub-type (is-A relationship) of an object in level 4 ( for example, polygon supertype). Queries can be managed so that a supertype returns all its subtypes, not vice versa.

FGIs actions are defined to be the description of the two geometries' interaction and their role in achieving the claimed function they belong to. FGIs actions are modelled as properties of the FGI edge. Functions are a group of FGIs that together execute the steps that perform the function; the order can also be specified. For example, a function to open a can have two steps represented in two FGIs: 1) a latch geometry presses a can cover geometry, then 2) the cover geometry lift and separates from the can body geometry. This is represented in graph notation as follows:

**latch –[press]→cover; cover-[separates]→can body**

Functions are modelled in a separate subgraph that is connected to the design root node and has unique function IDs property values that are referenced in all FGI edges participating in the completion of the function's steps. Therefore searching for a function will return the list of geometries involved in FGIs that are attributed to this function ID or name.

Behaviours are defined to be different behaviours for a function controlled by an attribute value. "For example, the radiator is always a source of heat to its environment, although this may be for either the purpose of heating this environment, or for cooling the fluid or coolant supplied to it, as for engine cooling". This can be modelled as a function of "changing the temperature of the environment" and two behaviours "heating" and "cooling". Therefore behaviour of a function is modelled as a sub function, by just using smart coding. If a function have only one behaviour, it can have an ID such as "f1". Another function with three different behaviours, would be given three different values as IDs: "f2_b1", "f2_b2", and "f2_b3".

The above model will create disjoint graphs for each patent, with edges to the product(s) it contains. Each product will have edges connecting it to the geometries it contains; geometries will have FGI relationships with other geometries with properties describing the function and behaviour it belongs to and the action it performs.

This model created the following pattern (expressed in Cypher Query) for all three levels that form the FAD model:
**match (p: patent {Patent_Number: "PatentUniqueIdentifier"})**
**optional match (p)-[:hasProduct]->(pr: product)**
**optional match (pr)-[:hasClaim]->(c)**
**optional match (pr)-[:hasGeometry]->(g1)**
**optional match (g1)-[fr:hasFGI]->(g2)**
**return p, pr, c, g1, fr, g2**

The above pattern describes the PatMine patent pattern. It is the same for the emerging design FAD, by replacing the node label from *patent* to *emergDesign* and using the CAD software filename as a unique identifier:

**match (p: emergDesign {filename: "cadSoftwareFileName"})**

With the following example constraints:

**CREATE CONSTRAINT ON (n:patent) ASSERT n.Patent_Number IS UNIQUE**
**CREATE CONSTRAINT ON (n:emergDesign) ASSERT n.filename IS UNIQUE**

Similar entity constraints can be applied to products, claims, and geometries. The referential constraints are imposed by matching before creating a link. The following example shows the pattern for FAD level two, the creation of geometries, after matching first to retrieve the encapsulating patent and product:

```
match (p:patent {Patent_Number: "PatentUniqueIdentifier"})
match (pr:product {Product_ID: "Product_ID"})
create (g:geometry {name: "Geometric_Feature_Name", Geometric_ID: "Geometric_Feature_ID",
PatMine_type: "Geometric_Feature_Type"})
create (pr)-[:hasGeometry]->(g)
return p, pr, g
```

Completing FAD level two, all FGIs need to be recorded. The following cypher statement is used to create an FGI relation between existing geometries after matching the patent/design and product:

```
match (p:patent {Patent_Number: "PatentUniqueIdentifier"})
match (pr:product {Product_ID: "Product_ID"})
match (g1:geometry  {name: "geomName1", Geometric_ID: "geomID1"})
match (g2:geometry  {name: "geomName2", Geometric_ID: "geomID2"})
create (g1)-[:hasFGI {Function_IDs: "Function_ID_List", action: "actionType"}]->(g2)
return p, pr, g1, g2
```

Retrieving all FAD level three that form the structure of one function, we can search all FGI relations that have Function_ID included in the function IDs list a property matching the search query:

```
match (g1)-[r1:hasFGI]->(g2) where "queryFunctionID" in r1.Function_IDs
match (p)-[:hasProduct]->(pr)
match (pr)-[:hasGeometry]->(g1)
return p, pr, g1, r1, g2
```

Once we have a FAD model for the emerging design, we can run a traditional full-text search with a query formed from the emerging design FAD keywords used over the FAD models of the PatMine patents database. These search results can be merged with similar text searches on the patents' original pdf documents. Full-text search can be achieved by using services such as Lucene/Solr or Elastic Search plug-ins for Neo4j or simply by searching in all property values:

```
match path = (p: patent)-[:hasProduct]->(pr: product)-[:hasGeometry]->(g1)-[fr:hasFGI]->(g2)
with nodes(path) AS x UNWIND x AS nodeLinks
with nodeLinks, [prop in keys(nodeLinks) WHERE nodeLinks[prop] in ["queryKeyword1", "queryKeyword2", …]] as matchRank
where size(matchRank) > 0
return nodeLinks, matchRank
```

The above path returns rows of paths from each patent in the Database down to each FGI it contains. These paths are scored for matched keywords, and the match score is then aggregated per patent. Ideas such as query expansion can be applied using the PatMine ontology or WordNet (Fellbaum, 1998) to use several queries for all synonyms and merge all results in the final list of relevant patents to the emerging design. Ranking these documents in descending order from the highest relevance can be as simple as counting the number of matched keywords (such as the size of matchRank variable in the previous query), and weighting the function to be higher weight than FGI action keywords, which can be higher than geometric features names and types. Similarly, using a synonym for a keyword in the query can produce a lower match rank than using the keywords required by the user or retrieved from the emerging design to be evaluated.

The semantic relationships between the keywords are already captured in the FAD model. Using this knowledge in guiding the search results retrieves higher-quality relevance. This will create queries for the three FAD levels. In FAD level one, queries are designed to search patents containing similar geometric features as the emerging design and come back with matchRank1. Then in FAD level two, queries are designed to search similar FGIs, i.e. containing the same geometries types and the same functional interaction. In FAD level three, queries are designed to search the function name or regular expressions, including it or any of its synonyms that match the function name in the emerging design and its structural components of FGIs action steps. FGIs comprising a function can be matched one-to-one to measure functional similarity or checked for conceptual similarity using higher levels terms in the PatMine embedded ontology (glossaries and synonyms lists). A final score is calculated by weighted sums of the three scores. The weights can be defined empirically as the Database is populated with more patents, and more emerging designs are scored by domain experts as similar or not. The following is a FAD level two semantic query, which searches similar FGIs using the PatMine conceptual geometric feature types and the kind of interactions between them.

**match (p)-[:hasProduct]->(pr)**
**match (pr)-[:hasGeometry]->(g1)**
**match (g1)-[r1:hasFGI]->(g2)**
**where g1.PatMine_TYPE =~ ".queryPatMineType1." and g2.PatMine_TYPE =~ ".queryPatMineType2."**
**and r1.action = "queryAction" and filter(funID IN r1.Function_IDs WHERE funID = "queryFunctionID")**
**return p, pr, g1, r1, g2, count(r1) as MatchRank2**

Using synonyms of function names, the where clause needs to be changed as follows:

**where r1.Function_Name in ["SynSet1"," SynSet2", …]**

The collaborative ontology-building approach is implemented simply now in PatMine SolidWorks Add-in © by loading values in combo-boxes user interface controls from CSV files pre-filled with the current PatMine ontology on loading of the tool. The designer is allowed to enter new values as required to enrich the ontology; then, the tool updates these CSV files on closing. Counters of usage are incremented every time a user selects an existing term or defines a new one. These counters are associated with the domain the annotated design belongs to. The more a value is used, the higher its counter and relevance to the domain will be. These counters represent votes (scores) it receives from users to quantify how a term value represents a similar concept to the related abstract type in the relevant domain. Studying these counters values after reasonably populating the Database with annotated patents, upper and lower ontologies can be deduced.

Representing the PatMine ontology along with the patent's extracted structured information as a graph model offered various benefits. First: the ontology/schema can evolve as more domain experts contribute more rules (constraints), entity types (nodes' labels), relationship types (edges labels), or characteristics for both (properties and their domain of values). Generalising the model to accept more rules from other domain experts is enabled naturally because of the schema-free nature of the unstructured neo4j graph database system. This feature enabled a collaborative ontology-building simple technique. Linking to other databases and knowledge bases will be a matter of connecting (adding edges) between existing nodes to the nodes in another graph structure extracted from these databases. An example is found in work in (Wang et al., 2017), which harvests XML metadata about a database and imports them to Neo4j database to identify various relationships.

Second: Since patents are semi-structured documents in which some information extraction can be automated within a particular domain, rules are defined for parsing the patent document to extract structured information from the unstructured text in different ways. Efforts to use NLP libraries such as Stanford Core NLP library (Bowman et al., 2016) and its Open information extraction (open IE) system (Angeli et al., 2016) have already started in the system; these libraries generally extract information in the form of relationships between entities, which is an intuitive graph model. This means that the collaborative ontology-building approach used currently, populated initially by the PatMine ontology glossaries and taxonomy decided by design engineering experts, can be complemented by automatic structured information extraction methods from the unstructured text found in the patents to be presented to the experts to accept and assign relevance scores or reject. This interactive learning technique allows the system to learn from previous decisions as more data is extracted, assigning it a more relevant score. There is also the Apache Unstructured Information Management Applications (UIMA) project to extract knowledge in the form of entities such as people, places, organisations, and their relationships from large volumes of unstructured texts, audio and video (Ferrucci and Lally, 2004). UIMA has been used to extract information from chemical patents (Jessop and Cambridge, 2011). Using existing lexicons such as WordNet (Fellbaum, 1998) to retrieve synonyms from the same language and/or to bridge translation barriers across different languages can enhance the conceptual modelling to higher levels than the wordings used in the patent document. The NLP open IE and/or UIMA extracted entities, and their interactions can be structured as a graph data model, while WordNet extends the search path to the synonyms of the extracted terms. Consequently, various levels of middle to lower ontology (more specifically for the various industrial domains and their glossaries, taxonomies and different relationships), will be refined and expanded using voting techniques among design engineers experts.

Third: performance is faster than sequential file access offered by Relational databases or separate files. The overview in (Vyawahare Harsha and P. P. Karde, 2015) mentioned a comparison of a three-level relationship (friends of friends) performing 150 times faster in graph modelling than in relational database modelling. A four-level deep query would perform better in graphs by a factor of 1000.

Fourth: Visualisation of graph data structures creates the FAD models automatically. Various visualisation libraries are discussed in the results section, and the suitable one is employed in the PatMine SolidWorks Add-in ©.

Fifth: representing the patents' or emerging design's extracted information (using a FAD model) in simpler geometric components and their interactions enables similarity quantification methods. Graph analytics methods can be used to summarise patents and identify industry trends.

# Results

Patent Mining (PatMine) © is developed as an add-in for Solid-Works using Visual Studio C# APIs. The tool is a front end that connects to the Neo4j database backend. The backend is a schema-free graph database; similarly, the front end was built to maintain this feature to an extent. Serialising the Neo4j query results required object types to capture the properties of the node. These objects are mainly five entities: 1) patent or emerging design, 2) product, 3) claims, 4) geometries, and 5) FGIs. The first label given to a node determines its main type. In this case, the basic node types are the said five main entities. Higher level abstractions are included as extra labels, such as geometries conceptual higher abstracted types. Column headers in the Excel sheets to upload to Neo4j are used as property names and their row contents as their values. This enables dynamic extraction of information as the knowledge of the patent document structure increases or as more domain experts to enrich the graph ontology with more object types, relationship types and taxonomy. Only extra properties can be added now to the main five entities and their relationships. It is straightforward to allow the user to define the main entities and relationships dynamically as future work. The tool adds a toolbar that contains buttons to invoke the functionalities offered by the system. Work is ongoing to enable the separation of PatMine and SolidWorks and to enable interfacing with other CAD tools as well. The following subsections explain the different functionalities offered by the tool.

### *FAD Annotation User Interface*

First: an interface to define the FAD model components for the emerging design. This is divided in three tabs: 1) the design and product names and the claims it provides; 2) the list of geometric features the design contains; and 3) the list of FGIs between the defined geometries. A corkscrew example is annotated in the UI screens and the visualisation and the SolidWorks design are illustrated in Fig 1.

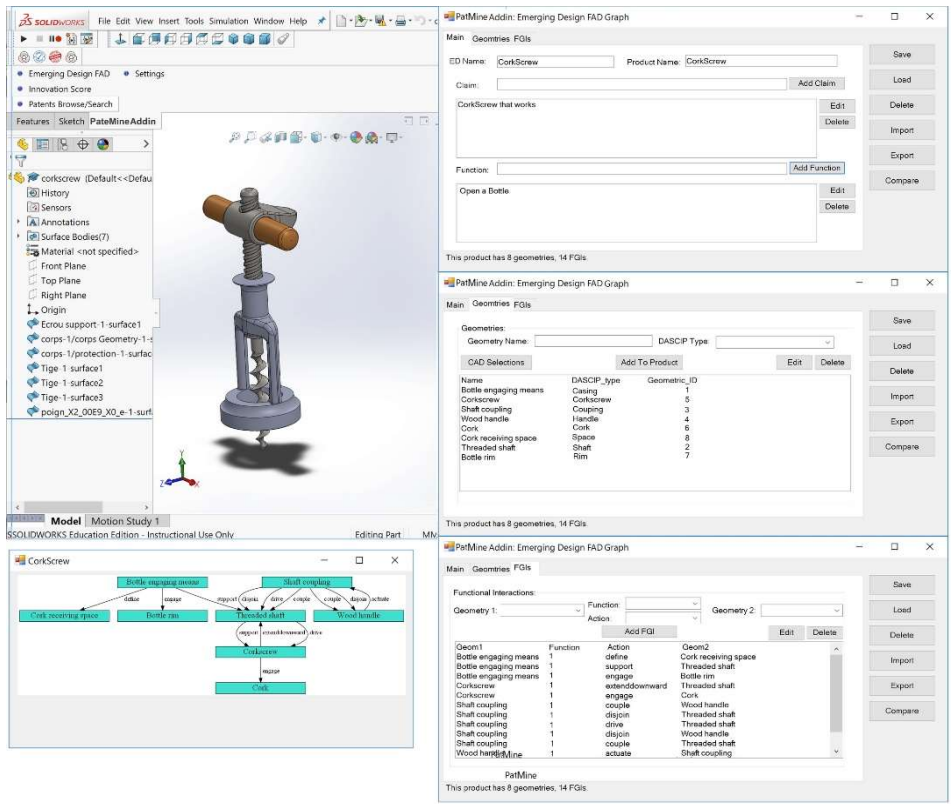

*Fig 1: PatMine SolidWorks Add-in © Corkscrew design FAD model Definition screens and visualisation*

## Patent Searching

The second functionality is to search and browse the FAD models stored in the PatMine patents database. Three types of search queries are currently implemented. The first is a semantic search query that accepts query keywords to search in the title, product, function, action and geometry nodes properties' values. A second full-text search is used to generate a dynamic query using keywords extracted from the FAD model values and can be changed by the user, as mentioned in the methods' last subsection. A third query option is to enable experienced users to write their own cypher statements. Fig 2 illustrates the three query options user interface. The resulting patents are displayed graphically with zooming functionality to investigate the interactions. Navigation buttons (First, Previous, Next, and Last) are used to move around between patents in the Database. Double-clicking the FAD image shows a thumbnail of the patent's FAD model graph stretched to be seen completely in a smaller area like the one shown in Fig 8. This user interface invokes the first user interface to update the FAD model for every patent in the Database to change its model contents, add and/or remove patents. Fig 3 shows the cork extracting apparatus patent displayed as a result set for an example search. When no values are entered in the search UI controls, all patents modelled in the DB are returned.

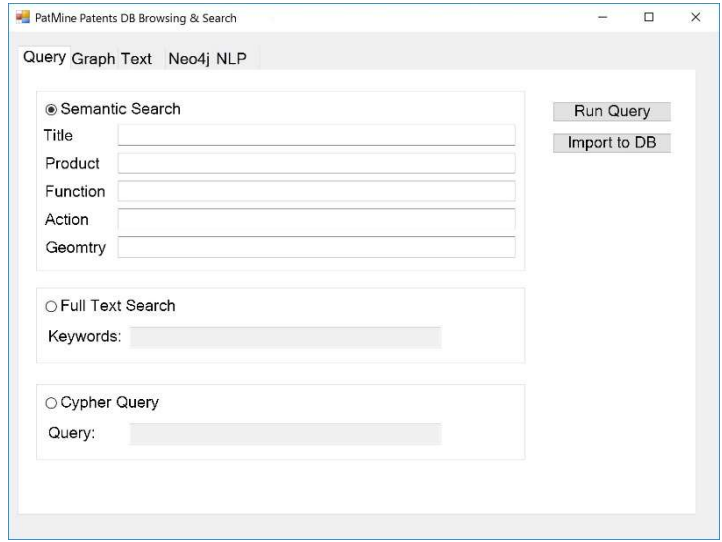

*Fig 2: PatMine SolidWorks Add-in © Patents Database Query Generation*

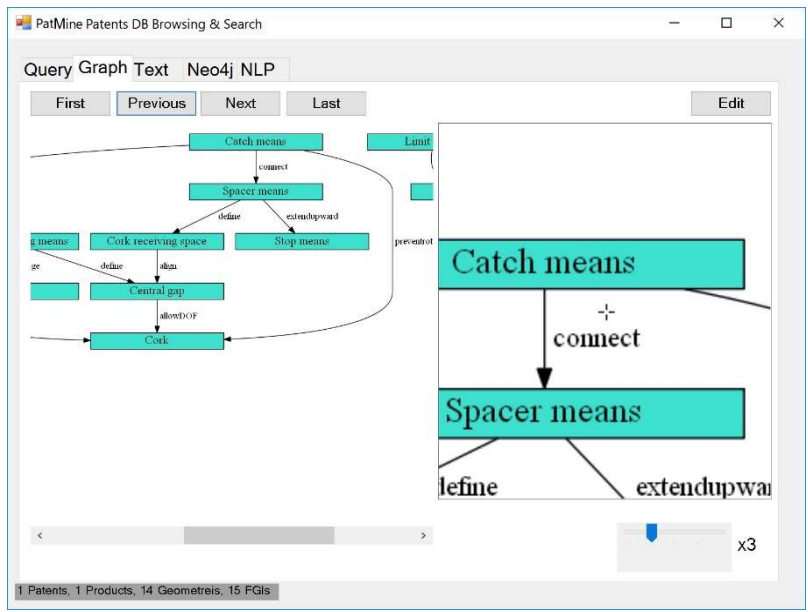

*Fig 3: PatMine SolidWorks Add-in © Patents Database Query Results Interface*

## Similarity Scoring

A third functionality is to compare the emerging design's FAD model with all patents' FAD models and display the comparison results. The comparison results are displayed to the user in terms of charts of matching scores, identifying overlapping geometries and interactions, corresponding patent document retrieval with the FAD annotations highlighted, and image thumbnails of the important drawing in a similar patent document. Simply in the FAD level one, similar geometric features are counted and given the smallest weight. Then, in FAD level two, the similar FGIs are counted and given a higher weight. Finally, in FAD level three, the combinations of FGIs forming the various functions are counted for overlaps, and the highest weight is given as a function of the number of FGIs in the matched function and a measure of the importance of the function by measuring its name similarity with the independent claim of the design. After populating the Database with patents' FAD models, another future experiment will implement and test the various scoring methods results and empirically decide the weights of each component to present to domain experts to evaluate the most appropriate method.

Current implementation uses an iterative search of the same geometric feature conceptual types between an emerging design and a patent identifies the count of the matching geometries, giving this count a weight of 10. Another iterative search of overlapping FGI (geometry –[action]->geometry) relationships, and a resulting count is produced, given a weight of 20. A third iterative search of functions (groups of FGIs that implement a function) produces a count of overlapping functions, and multiplied by a weight of 30. A match rank between an emerging design and a given patent is calculated as a weighted sum of the above three counts as follows:

**_Match Rank = ((geometries count * 10) + (FGI count * 20) + (Functions count * 30)) / 60_**

These weights are assumed based on the complexity level of each component. More advanced methods can be used to calculate these weights after manually annotating and scoring more patents/designs, such as perceptrons. A normalisation step across all patents match scores is done using the following equation:

**_Match Rank = (Match Rank – Minimum Match Rank ) / (Maximum Match Rank – Minimum Match Rank )._**

The illustration in Fig 4 shows some screens of the scoring results for the corkscrew example, showing a score with only one count. Double clicking the chart bar, displays the overlapping FAD annotation (in this example two geometries and one Functional interaction between them. Retrieving the original patent document is achieved by single clicking on the corresponding bar. To make the similarity scoring conceptual, the ontology type of the geometry is used for matching rather than the name given by the user. More advanced techniques can be used to allow for indirect relationships such as network alignment algorithms. The term alignment here is used to simply refer to the identification of the overlapping region between two FAD models graph data structure.

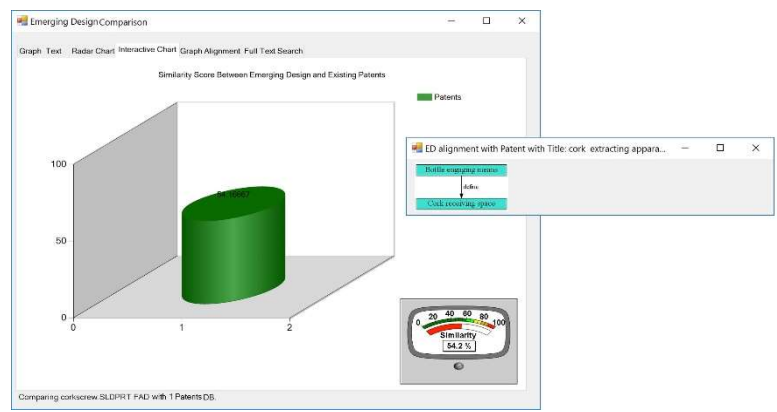

*Fig 4: PatMine SolidWorks Add-in © Corkscrew example design scored against a 54% similar Patent, and overlapping interaction identified*

## *Visualisation*

Data visualisation is an important aspect of exploratory data analysis. Visualisation is a sort of summarisation of information that otherwise requires much more effort to read from large amount of data. Patent documents and the semi-structured information they contain, are visualised as graphs connecting patents citations such as the study in (Huang et al., 2003). Citation graphs contain much less patent's inner information than network-based patent analysis methods, in which the overall relationship among patents is represented as a visual network such as the work in (Yoon and Park, 2004). The unstructured information in the patent documents contain deeper insights about the patents' contents and their relevance. The work in (B.-U. Yoon et al., 2002; H. Atsushi and T. YUKAWA, 2004, 2004; S. Lee et al., 2009; Yoon and Park, 2004) presented patent maps in various approaches such as technology vacuum map, claim point map, and technology portfolio map.

Graph data models enable straight forward visualisation of relationships and patterns. Graphical interfaces were first introduced to support the ER model as schema management such as schema definition, schema browsing, and query formulation. Examples include: GUIDE, DDEW, SKI, ISIS, and SNAP. ISIS and LID allow also for data browsing and relationships traversal to find related objects. SKI, ISIS, and SNAP generate graph-like visualisation of the schema (Hull and King, 1987). Graph drawing is a branch of computational geometry with various algorithms and systems available in the literature. These systems account for topological and geometric concepts such as planarity testing and embedding, crossings and planarisation, symmetric drawings, and proximity drawings. These systems offer various drawing options such as tree and hierarchical drawing, spine/radial/circular drawing, rectangular drawing, simultaneous embedding drawing and force-directed drawing among others (Tamassia, 2013). One of the earliest graph visualisation systems is Hy+ (Consens and Mendelzon, 1993).

The graph formats: Graph Mark-up Language (GraphML) is a graph data structure based on XML technology useful for integrating data elements in web documents. Combined with Scalable Vector Graphics (SVG). GraphML is useful for web drawings of graph data structures, such as D3.js. This library is the foundation on which the default Neo4j Server browser uses as a powerful, customisable data visualisation tool using Graph-Style-Sheet (GRASS) for styling. PatMine SolidWorks Add-in © uses Neo4j Server browser in the backend and is making it available online for other users in case they know cypher to query and interact with the PatMine SolidWorks Add-in © patents graph database in an available public database instance. Neo4j Server browser employs a simple force-directed graph drawing algorithm that uses character co-occurrence as the force guiding a physical simulation of charged particles and springs. This algorithm places related characters taken, for example, from node labels and properties' values in closer proximity while unrelated characters are farther apart. An Example corkscrew design and Cork–screw apparatus Patent as visualised in the Neo4j Server Browser is shown in Fig 5 and Fig 6, respectively.

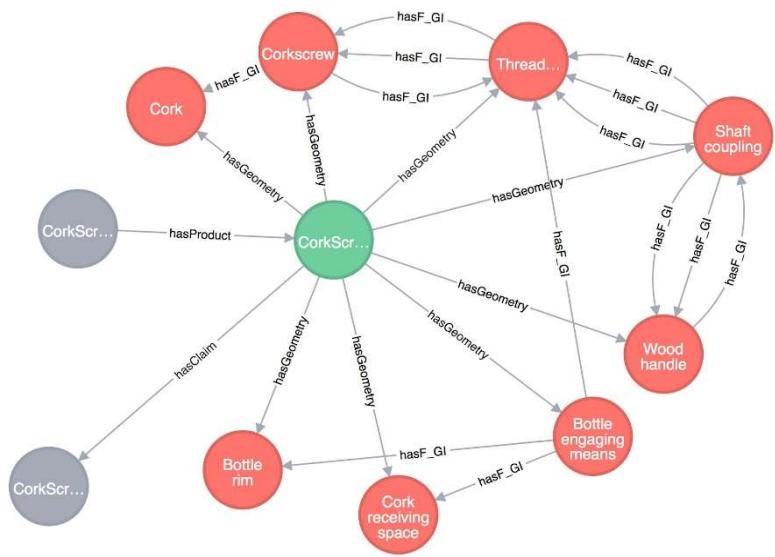

*Fig 5: Corkscrew Design FAD visualisation in Neo4j Server Browser*

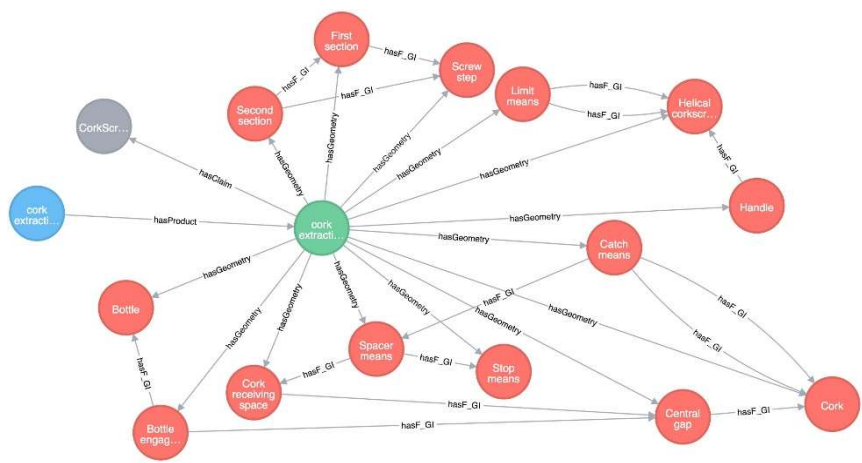

*Fig 6: Corkscrew Apparatus Patent FAD visualisation in Neo4j Server Browser*

There are other tools to achieve customised graph visualisation. Alchemy.js is a visualisation for graph data structures such as Neo4j databases that uses GraphJSON format, which is similar to GeoJSON. Other visualisation tools that don't offer community free editions are Linkurious.js Visualization Library and KeyLines, among other solutions.

A more formal FAD visualisation is developed in the software to highlight the geometric interactions for a given product. The front end of the PatMine SolidWorks Add-in © required a library to integrate with SolidWorks APIs that are available in .Net C# and Visual Basic. The following options for graph drawing are investigated. The Open Graph Drawing Framework (OGDF) is a C++ library of algorithms and data structures for graph drawing based on the LEDA library. There are in-memory graph tools characterised by their restriction to work with small graphs. For example, complex analysis tools (e.g., Cytoscape), and graph visualisation tools (e.g., JUNG, IGraph, Gephi, NodeXL and GraphViz).

The .Net wrapper for the GraphViz suite of tools was used (Emden R. Gansner and Stephen C. North, 2000), which is based on the dot language (Emden Gansner et al., 2002). This required writing the code for transforming the Neo4j query result tabular format, which repeats graph contents in all resulting rows, to their serialised Object oriented entities and relationships, then to the equivalent dot language format to produce the required visualisation. Abstracting this code can produce generic graph structure file format converters across different file formats to use features from the different libraries and to enable various graph analytics algorithms. GraphViz is based on the Sugiyama method of drawing directed graphs, which separates the nodes into layers. The corkscrew design vs the corkscrew apparatus patent examples used above is visualised in PatMine SolidWorks Add-in © screens in Fig 7 and Fig 8, respectively. This illustration is useful for identifying clusters of interactions that could imply one or more function(s), geometries that are not interacting with others, and functions that are divided into clear sub-functions because they are producing different subgraphs.

The generated dot language abstracted the patents/designs and product nodes and used only geometries as nodes with the FGIs among them. The converter enables the use of different abstraction levels. This can abstract the geometrical names used by the designer to their PatMine abstract types or any higher level of abstraction to produce graph representations on any abstraction level for the same design.

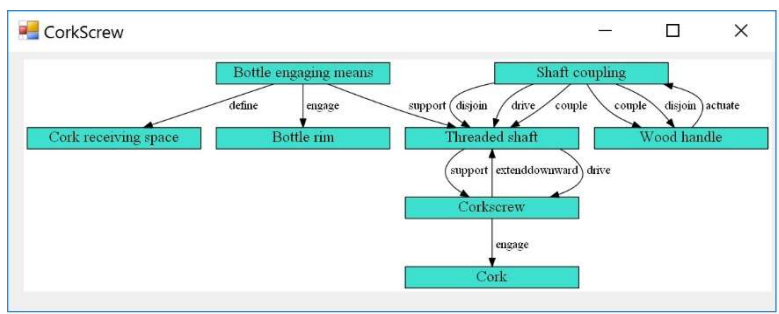

*Fig 7: Corkscrew FAD visualisation in PatMine SolidWorks Add-in ©*

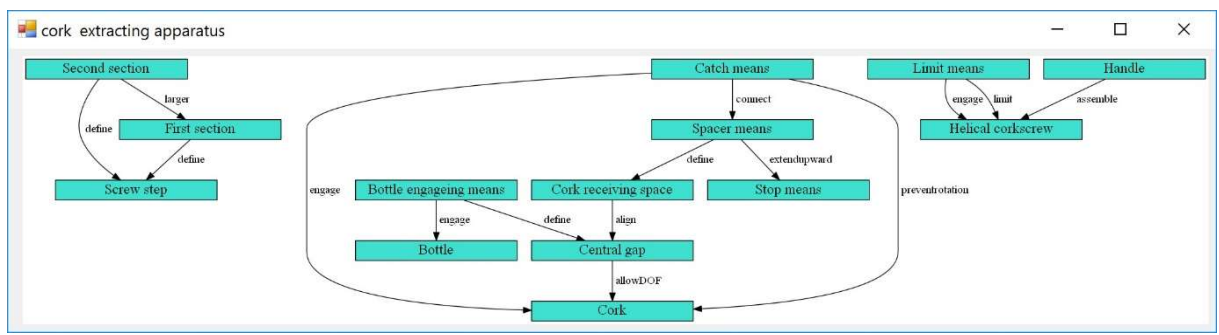

*Fig 8: corkscrew apparatus patent FAD visualisation in PatMine SolidWorks Add-in ©*

A fourth functionality exposes all environment variables and system settings to be updated by the user to enable dynamic behaviour of the system as much as possible, such as file path to patent documents, and path to dot engine. The Database is bulk populated from Excel sheets describing the FAD annotations uploaded directly to the Database back end. Direct loading of CSV into the Neo4j server using cypher statements is performed sometimes or through the PatMine SolidWorks Add-in © User Interface. The Neo4j database backend is available on a single instance server, to which users can connect to and upload their FAD modelling of patents and emerging designs. An interface to change the ontology taxonomy, glossaries, and relationships will be integrated soon to enable voting for the best way to model the various domains. If the Database grows, distributed deployment of Neo4j on a reasonably sized cluster will be investigated. These future steps will enable a collaborative ontology-building approach to widen our understanding of the upper ontology across all design domains and a lower and more specific ontology for subdomains.

# Conclusion

The presented work contributes a novel method of modelling both an ontology and a semantic database using a graph data model. Employing the current system as a basis for more AI and machine learning research to build semantic networks using inference and logic techniques such as unification, lambda calculus, and Peirce's existential graphs will enrich the PatMine patents knowledge-base and its analytics (Sowa, 1992). The merging of the taxonomy, glossaries, and relationships' definitions, and the storage of the entities using a schema-free model enabled collaborative ontology building as more knowledge is acquired. It also allowed the efficient storage of entities based on their relationships for faster retrieval of deeper queries. Furthermore, the visualisation of the FAD models was automated using this efficient and intuitive data model. More quantitative scoring methods are enabled by using this model that is graphically justified, i.e. designers will see the overlapping nodes highlighted in the FAD-generated image. These scoring methods can identify how innovative a new emerging design is compared to patents in the PatMine database. Another study is planned to compare various similarity scoring methods from the literature on top of this design and evaluate their suitability to engineering designs' functional similarity scoring objectives. This will alert designers to areas in their design of concern to focus on to avoid the overlap with existing patents and possible infringement. Another benefit of graph modelling the functional analysis of engineering designs in patents is that patents can be checked for similarity against each other, and industry trends can be statistically identified and visually summarised by various attributes such as location (country), time (year), industrial domain, inventive principles, geometric objects and functional interactions, among other attributes extracted from patents.

Future plans and directions are briefly discussed in the following. Extracting the FAD model is entered into the PatMine SolidWorks Add-in © User Interface, explained in section seven, by designers and domain experts to model the patent document or the emerging designs or filled-in template Excel sheets to be uploaded to the backend. This limitation requires extra input from the designer to annotate existing patents and to annotate the emerging designs of interest to check their similarity. Since the annotation level of detail will differ from one expert to another, the similarity scoring will be based on the level of detail the annotators have included in the FAD modelling step and label it a subjective score rather than pure objective as would be if full automation in the annotation step is achieved.

Since the information extraction approach required to build the FAD model is very systematic, then given a patent document, this operation can be automated using Natural Language Processing (NLP) libraries. When modelling an emerging design, the current interface with SolidWorks captures the currently selected faces in the open design document as geometries to add automatically to the design being annotated. Further development can lay out the annotation over the design drawing in SolidWorks.

Further development of the tool will focus on developing the ontology to guide the user through the abstracted types for the various selected faces in the CAD document and the various possible functional interactions for the selected types. This will limit the designers' possible mistakes of defining an impossible interaction. Separating the PatMine system from the SolidWorks interface, including various engineering drawing applications' interfaces such as AutoCAD, and generalising the geometries selections, drawing and annotating the CAD document will widen the use of the proposed system.

The current status of PatMine SolidWorks Add-in © is that it accepts extra properties and their values to the main five entities introduced earlier as 1) patent or emerging design, 2) product, 3) claims, 4) geometries, and 5) FGIs. These extra properties are supplied freely by the user in Excel sheets after the main required labels and properties. To enable genuine schema-free design in the front end as well as it is genuine in the backend, future work will focus on removing the main entities defined. Only one generic node type and one generic relationship type can be used. Both can accept any number of labels and properties for hierarchical abstractions and detailed descriptions based on what is available in the unstructured information extracted freely from the patents' documents. The user can then define a dynamic schema to force the model to abide by its constraints or accept the schema-free design and use the dynamic query generation interface carefully.

The work presented in this study will facilitate building a web crawler to search the online patents databases to automatically extract their FAD models using NLP and upload them to a distributed database for semantic queries and similarity scoring, among many other graph analytics of interest. The tool will soon be available for download connecting to the Neo4j server that will be installed in (neo4j.manalhelal.com). If the number of concurrent users increases and the amount of annotated patents becomes big, a single-instance solution might not be very useful. A cluster of servers for the database backend will be required. Graph databases can perform less efficiently on a low-cost cluster. Neo4j uses a master-slave approach to synchronise changes over slave nodes. If Neo4j does not scale well with the expected growth and available cluster configurations, other more efficient distributed graph database systems options will be investigated to plan a migration plan. InfiniteGraph supports large-scale graphs in a distributed environment. There is also Giraph, which is a graph processing infrastructure that runs on Hadoop, CloudGraph, GoldenOrb, and Phoebus, among other solutions. Although NoSQL and graph models seem like a solution to Big Data and semantic modelling requirements, the no-schema approach came at the cost of losing the ACID (Atomicity, Consistency, Isolation, and Durability) transactions benefits of the relational models. Recently Google announced Spanner distributed relational Database with ACID transactions, availability and scalability of the NoSQL systems.

A wiki for the project is maintained gradually at http://patmine.manalhelal.com to enable interested parties to follow up with the ongoing development efforts and contact the developer for suggestions and feedback.